\documentclass{aa}
\pdfoutput=1
\usepackage[utf8]{inputenc}
\usepackage[varg]{txfonts}
\usepackage{natbib}
\usepackage{caption}
\usepackage{rotating}
\usepackage{subcaption}
\usepackage{xcolor}
\usepackage{afterpage}
\usepackage{booktabs}
\usepackage{chemformula}

\usepackage[markup=bfit]{changes}
\usepackage{xcolor}
\usepackage{mathtools}
\usepackage{xspace}
\usepackage{soul}
\usepackage{cancel}
\usepackage{float}

\usepackage[separate-uncertainty = true,multi-part-units=single, group-digits = integer, group-four-digits = true,per-mode=reciprocal]{siunitx}

\usepackage{natbib,twoopt}
\usepackage[breaklinks]{hyperref}      
\bibpunct{(}{)}{;}{a}{}{,}             
\definecolor{cobalt}{rgb}{0.06, 0.2, 0.65}
\hypersetup{
  colorlinks,
  citecolor=cobalt,
  linkcolor=[rgb]{0.8, 0.2, 1.0},
  urlcolor=cobalt,
}
\makeatletter
  \newcommandtwoopt{\citeads}[3][][]{\href{http://adsabs.harvard.edu/abs/#3}%
    {\def\hyper@linkstart##1##2{}%
     \let\hyper@linkend\@empty\citealp[#1][#2]{#3}}}
  \newcommandtwoopt{\citepads}[3][][]{\href{http://adsabs.harvard.edu/abs/#3}%
    {\def\hyper@linkstart##1##2{}%
     \let\hyper@linkend\@empty\citep[#1][#2]{#3}}}
  \newcommandtwoopt{\citetads}[3][][]{\href{http://adsabs.harvard.edu/abs/#3}%
    {\def\hyper@linkstart##1##2{}%
     \let\hyper@linkend\@empty\citet[#1][#2]{#3}}}
  \newcommandtwoopt{\citeyearads}[3][][]%
    {\href{http://adsabs.harvard.edu/abs/#3}
    {\def\hyper@linkstart##1##2{}%
     \let\hyper@linkend\@empty\citeyear[#1][#2]{#3}}}
\makeatother

\usepackage[separate-uncertainty = true,multi-part-units=single,group-separator={,}, group-digits = integer, group-four-digits = true,per-mode=reciprocal]{siunitx}
\usepackage{xspace}

\newcommand{\numpm}[3]{$#1^{#2}_{#3}$}

\newcommand{\kpvsys}{K$_p$-V$_{\rm sys}$\xspace}
\newcommand{\metric}{$\Delta_{\text{Ti} - \text{Fe}}$\xspace}

\newcommand{\fedetections}{WASP-76~b, KELT-20~b, WASP-121~b, WASP-178~b, TOI-1518~b, HAT-P-70~b and WASP-189~b\xspace}
\newcommand{\tidetections}{WASP-121~b, TOI-1518~b, HAT-P-70~b and WASP-189~b\xspace}

\usepackage{afterpage}
\usepackage[markup=bfit]{changes}
\usepackage{xcolor}

\begin{document}
\title{A population view of transiting hot giant exoplanets: \\ Tracing Fe and Ti chemistry with ESPRESSO and MAROON-X}

\author{Bibiana~Prinoth\inst{1,2}\thanks{ESO Fellow}
\and
Vivien~Parmentier\inst{3} 
\and
Stefan~Pelletier\inst{4}
\and
Daniel~Kitzmann\inst{5}
\and
Julia~V.~Seidel\inst{3, 6}\thanks{Poincaré Fellow}
\and 
Adrien~Simonnin \inst{2}
\and
Valentin~De Lia \inst{3}
\and
Sydney~Vach \inst{1, 7}
\and
Jens~Kammerer\inst{1}$^\star$
\and
Matteo~Brogi\inst{8,9}
\and
Florian~Debras\inst{10}
\and
Michael~R.~Line\inst{11}
}

\offprints{B. Prinoth, \\ \textbf{ \email{bibiana.prinoth@eso.org} }}

\institute{
European Southern Observatory, Karl-Schwarzschild-Strasse 2, 85748 Garching, Germany
\and 
Lund Observatory, Division of Astrophysics, Department of Physics, Lund University, Box 118, 221 00 Lund, Sweden 
\and 
Laboratoire Lagrange, Observatoire de la Côte d’Azur, CNRS, Université Côte d’Azur, Nice, France
\and
Observatoire astronomique de l'Universit\'{e} de Gen\`{e}ve, 51 chemin Pegasi 1290 Versoix, Switzerland 
\and
University of Bern, Physics Institute, Division of Space Research \& Planetary Sciences, Gesellschaftsstr. 6, 3012, Bern, Switzerland
\and
European Southern Observatory, Alonso de C\'ordova 3107, Vitacura, Regi\'on Metropolitana, Chile 
\and
Centre for Astrophysics, University of Southern Queensland, Toowoomba, QLD 4350, Australia 
\and
Department of Physics, University of Turin, Via Pietro Giuria 1, I-10125 Torino, Italy 
\and
INAF – Osservatorio Astrofisico di Torino, Via Osservatorio 20, I-10025 Pino Torinese, Italy 
\and
IRAP, Université de Toulouse, CNRS UMR 5277, 31400 Toulouse, France 
\and 
School of Earth and Space Exploration, Arizona State University, 781 Terrace Mall, Tempe, AZ, 85287, USA 
}

\date{Received: April 1, 2026 -- Accepted: June 07, 2026}

\abstract{
Hot and ultra-hot Jupiters offer a unique laboratory to study atmospheric chemistry at the population level using ground-based high-resolution spectroscopy. Iron (Fe) and titanium (Ti) are key tracers of thermal and chemical structure, yet they exhibit markedly different observational trends across the population. We present a homogeneous re-analysis of high-resolution transmission spectra for ten hot and ultra-hot Jupiters observed with ESPRESSO on ESO’s Very Large Telescope and MAROON-X on Gemini-North. 
We search for neutral Fe and Ti absorption and perform injection–recovery tests using forward models spanning a range of Ti-depletion levels and temperature–pressure profiles.
To enable direct comparison across planets observed with different instruments and signal-to-noise ratios, we introduce the relative cross-correlation metric, $\Delta_{\text{Ti} - \text{Fe}}$. We detect Fe in seven planets and Ti in four planets at significances above $5\sigma$. Across the population, $\Delta_{\text{Ti} - \text{Fe}}$ decreases sharply towards lower equilibrium temperatures. Under the assumption of equal Ti depletion across planets, isothermal models fail to reproduce this trend, instead requiring a temperature-dependent depletion of Ti that increases toward cooler planets, consistent with cold-trapping processes in cooler atmospheres. Models with inverted temperature–pressure profiles naturally reproduce the decline without invoking such temperature-dependent depletion. In these atmospheres, Ti is converted into TiO in deeper, cooler layers and subsequently removed from the gas phase through condensation into Ti-bearing species, leading to a strong suppression of the observable atomic Ti signal while Fe remains largely atomic. Nevertheless, even in these gradient models, an additional overall depletion of Ti relative to Fe is required to match the hottest planets. Our results demonstrate that the observable refractory chemistry is governed by the interplay of molecular partitioning, ionisation, condensation, and cold-trapping processes, as well as the vertical structure of ultra-hot Jupiter atmospheres. Although these results reveal clear population-level trends, additional observations will be necessary to distinguish between temperature-dependent cold-trapping and overall depletion scenarios. Expanding homogeneous high-resolution surveys to larger and more diverse samples, targeting both emission and transmission observations, will refine these constraints and provide critical insights into the chemistry of strongly irradiated giant planets.
}


\titlerunning{Tracing Fe and Ti chemistry with ESPRESSO and MAROON-X}

\maketitle

\section{Introduction}

Ultra-hot and hot Jupiters have long served as workhorses for atmospheric characterisation with ground-based high-resolution spectroscopy \citep[e.g.,][]{snellen_orbital_2010, brogi_signature_2012, birkby_detection_2013, hoeijmakers_atomic_2018, prinoth_titanium_2022, pelletier_vanadium_2023, seidel_vertical_2025}, thanks to their inflated radii, and short orbital periods. Observed in both transmission and emission, their atmospheres probe extreme physical conditions and provide unique insights into atmospheric chemistry and dynamics \citep[e.g.,][]{ehrenreich_nightside_2020, line_solar_2021, pelletier_where_2021, brogi_roasting_2023, Zhang2026}.

Optical high-resolution spectrographs ($\mathcal{R} \gtrsim \num{70000}$) have enabled the detection of numerous refractory metals in these atmospheres, most notably iron and titanium after their first discovery in ultra-hot Jupiter atmospheres \citep{hoeijmakers_atomic_2018}. These detections have primarily relied on the cross-correlation technique \citep[e.g.][]{snellen_orbital_2010}, which exploits the combined signal of hundreds to thousands of absorption lines to enhance weak planetary signatures above stellar and telluric contamination \citep[see][for recent reviews]{birkby_exoplanet_2018, snellen_exoplanet_2025}. Applied systematically across multiple planets, this method allows comparative studies that reveal how physical and chemical processes shape observable atmospheric composition.

Observational studies have shown that iron is nearly ubiquitous in ultra-hot Jupiter atmospheres ($T_{\rm eq} \gtrsim \SI{2200}{\kelvin}$). Both neutral and ionised iron have been detected across a broad range of equilibrium temperatures in transmission and emission observations \citep[e.g.,][]{borsa_gaps_2019, casasayas-barris_atmospheric_2019, ben-yami_neutral_2020, gibson_detection_2020, hoeijmakers_high-resolution_2020, nugroho_searching_2020, stangret_detection_2020, yan_temperature_2020, cabot_toi-1518b_2021, cont_detection_2021, kesseli_confirmation_2021, Kasper2021, BelloArufe_mining_2022, bello-arufe_exoplanet_2022, gandhi_spatially_2022, Herman2022, kesseli_atomic_2022, PaiAsnodkar2022, Pino_2022, Yan2022, borsato_small_2023, johnson_pepsi_2023, Lowson2023, ramkumar_high-resolution_2023, lesjak_retrieving_2024, petz_pepsi_2024, silva_espresso_2024, Stangret2024, young_searching_2024, Bazinet2025, basinger_pepsi_2025, householder_KPF_2025, Ramkumar2025, vaulato_atmospheric_2025, Vaulato_hydride_2025, Sanchez2025, vanSluijs2025, Yang2025, Lenhart_pepsi_2026, Bonidie_2026}. Titanium, by contrast, has proven more difficult to detect. While strong neutral and ionised Ti signatures are observed in some of the most strongly irradiated planets \citep[e.g.,][]{hoeijmakers_atomic_2018, cont_atmospheric_2022, prinoth_titanium_2022, stangret_high-resolution_2022, borsato_mantis_2023, jiang_detection_2023, scandariato_pepsi_2023, guo_detection_2024, darpa_gaps_2024, simonnin_time_2024, prinoth_titanium_2025, Zhang2026}, many cooler objects, despite having clear iron detections, exhibit weak or entirely absent Ti \citep[e.g.,][]{hoeijmakers_hot_2020, merritt_non-detection_2020, gandhi_retrieval_2023, maguire_high-resolution_2023, pelletier_vanadium_2023}. Additionally, there appears to be a paucity of TiO detections, with one robust ground-based detection confirmed across multiple spectrographs in WASP-189~b \citep{prinoth_titanium_2022, prinoth_time-resolved_2023}, and several other tentative detections awaiting confirmation \citep[e.g.,][]{nugroho_high-resolution_2017, sedaghati_detection_2017}. 
Meanwhile, lower resolving power observations with space-based observatories such as HST and JWST can struggle to unambiguously resolve spectral bands of TiO \citep[e.g.,][]{evans_detection_2016, Pelletier_2026}.

Detailed studies performing atmospheric retrievals and injection-recovery analyses of individual systems consistently find titanium to be depleted relative to iron. For example, the atmosphere of WASP-121~b \citep[$T_{{\rm eq}} \approx \SI{2350}{\kelvin}$, ][]{delrez_wasp-121_2016} exhibits significantly weaker Ti absorption compared to Fe across multiple independent studies \citep[e.g.,][]{gibson_relative_2022, maguire_high-resolution_2023, pelletier_crires_2024, hoeijmakers_mantis_2024, prinoth_titanium_2025}. At somewhat lower equilibrium temperatures, Ti appears to vanish entirely from the observable atmosphere in planets such as WASP-76~b  \citep[$T_{{\rm eq}} \approx \SI{2200}{\kelvin}$,][]{ehrenreich_nightside_2020}, where the abundance of iron and other refractory metals are consistent with stellar while Ti is comparatively significantly depleted \citep{gandhi_retrieval_2023, pelletier_vanadium_2023}. Similarly, with higher equilibrium temperatures, HAT-P-70~b, TOI-1518~b, and WASP-189~b, also exhibit depleted Ti relative to Fe, however, less so compared to cooler systems \citep{simonnin_time_2024, gandhi_retrieval_2023}.

Several mechanisms have been proposed to explain this observed trend of titanium signals that are weaker than predicted by models. One possibility is that the elemental abundance of Ti relative to Fe differs from that of the host star, for example as a result of formation processes in the protoplanetary disc. However, this scenario is generally considered unlikely, as both elements are refractory and are expected to be incorporated into planets primarily in solid form during formation, thereby largely preserving their relative abundances \citep[e.g.,][]{lothringer_new_2021}.

Another possibility is that titanium is preferentially removed from the atomic phase through recombination into molecules such as TiO. In this case, Ti may remain present in molecular form but becomes undetectable in cross-correlation analyses using atomic Ti. However, this explanation is also challenged by low-resolution observations, which indicate that TiO itself can be significantly depleted in some ultra-hot Jupiter atmospheres \citep[e.g.,][]{evans_optical_2018, Pelletier_2026}, suggesting that additional processes must be involved that deplete the overall abundance of titanium-bearing species.

A third possible explanation is the operation of a cold-trap mechanism that removes titanium from the gas phase of the observable atmosphere through condensation and rainout processes \citep[e.g.,][]{evans_optical_2018, hoeijmakers_hot_2020}. Titanium readily forms TiO and TiO$_2$ molecules, which can subsequently condense into refractory compounds such as perovskite (\ch{CaTiO3}) and other titanium oxides \citep{lodders_titanium_2002, Kitzmann_2023}. These condensates form even at relatively high temperatures, where Ti-species are expected to condense at approximately \SI{1600}{\kelvin}, compared to about \SI{1400}{\kelvin} for iron \citep{lodders_solar_2003}. In the intensely irradiated atmospheres of ultra-hot Jupiters, TiO is largely thermally dissociated on the dayside, while on the cooler nightside or in deeper atmospheric layers it can recombine and subsequently condense \citep{spiegel_can_2009, parmentier_3d_2013}. Three-dimensional circulation models predict that these condensates can rain down towards the deep atmospheric layers fast enough to prevent efficient vertical mixing and the replenishment of titanium in the observable upper atmosphere \citep{Parmentier_2016}.  In theory, such a process could deplete titanium more efficiently than iron. However, given that the \SI{200}{\kelvin} difference between the condensation curves of Ti and Fe is small compared to the $\sim$\SI{1000}{\kelvin} day-to-night contrast in these planets \citep[e.g.][]{Splinter_2025}, both species are expected to condensed together and form mixed, dirty grains~\citep{Helling_2019}. 

While TiO$_2$ has a relatively low self-condensation energy barrier~\citep{lee_dust_2015} that can allow for efficient heterogeneous nucleation~\citep{fegley_atmospheric_1996}, the efficiency of this mechanism to preferentially remove Ti compared to Fe thus still needs to be demonstrated.

Taken together, all these studies suggest that titanium depletion is a common feature of ultra-hot Jupiter atmospheres, but its physical origin remains to be confirmed. Disentangling the relative roles of molecular partitioning, ionisation, condensation, and cold trapping, therefore, requires systematic observational constraints across a wide range of planetary conditions. In particular, identifying trends with parameters such as equilibrium temperature, surface gravity, composition, and stellar irradiation may help distinguish between competing mechanisms. While historically most high-resolution studies have focused on individual benchmark planets, the field is now increasingly shifting toward homogeneous population-level analyses designed to identify such trends \citep[e.g.,][]{langeveld_survey_2022, stangret_high-resolution_2022, gandhi_retrieval_2023, lothringer_library_2025}. Such studies are essential for separating genuine observational trends from observational biases introduced by different instruments, analysis techniques, or data quality.

In this work, we present a homogeneous population study of iron and titanium absorption in the transmission spectra of ten hot and ultra-hot Jupiters observed with the ESPRESSO spectrograph on ESO’s Very Large Telescope and the MAROON-X spectrograph on Gemini North. We reanalyse all datasets using a consistent methodology and quantify the relative trend of Fe and Ti by comparing the observed population trends to forward-model predictions.

\section{Observations and sample selection}
\label{sec:observations}

Our study focuses on hot and ultra-hot Jupiters with high-resolution optical spectra obtained using either MAROON-X at Gemini-North \citep{seifahrt_maroon-x_2018} or ESPRESSO at ESO's Very Large Telescope at Paranal Observatory \citep{pepe_espressovlt_2021}. These instruments were chosen for their combination of high spectral resolution ($R \gtrsim$ \num{85000}), broad wavelength coverage (ESPRESSO: 378 -- 789~nm, MAROON-X: 500 -- 920~nm) in the optical where iron and titanium are absorbing strongly, and their installation on 8\,m class telescopes allowing for sufficiently high signal-to-noise ratio (S/N) with relatively short exposures, thus limiting smearing of the exoplanet signal due to its rapidly changing line-of-sight velocity. This allows for detection of weak metal absorption even in atmospheres with depleted abundances \citep[e.g.][]{prinoth_titanium_2025}. 

From the population of known gas-giant exoplanets ($0.75 \leq$ R$_{\rm p}$ [R$_{\rm Jup}$] $\leq 2.5$), we identified all systems observed in transmission with MAROON-X or ESPRESSO. In total, 16 planets match these criteria. Several systems were excluded from the final sample: KELT-4~A~b, MASCARA-4~b, and KELT-7~b due to strong stellar pulsations \citep{eastman_kelt-4ab_2016, stangret_high-resolution_2022, zhang_transmission_2022}, which significantly distort the cross-correlation signals; WASP-19~b due to low signal-to-noise observations dominated by instrumental systematics \citep{sedaghati_spectral_2021}; KELT-17~b due to the peculiar chemical composition of its host star, which has been shown to hinder the detection of even iron \citep[see e.g.][and study in prep.]{stangret_high-resolution_2022}; and KELT-9~b due to its extreme equilibrium temperature, which places it in a substantially different physical and chemical regime characterised by strong ionisation \citep[e.g.][]{hoeijmakers_atomic_2018, Kitzmann_2018}. The resulting sample used for our analysis therefore comprises 10 hot and ultra-hot Jupiters.  

\begin{figure}[t]
    \centering
    \includegraphics[width=\linewidth]{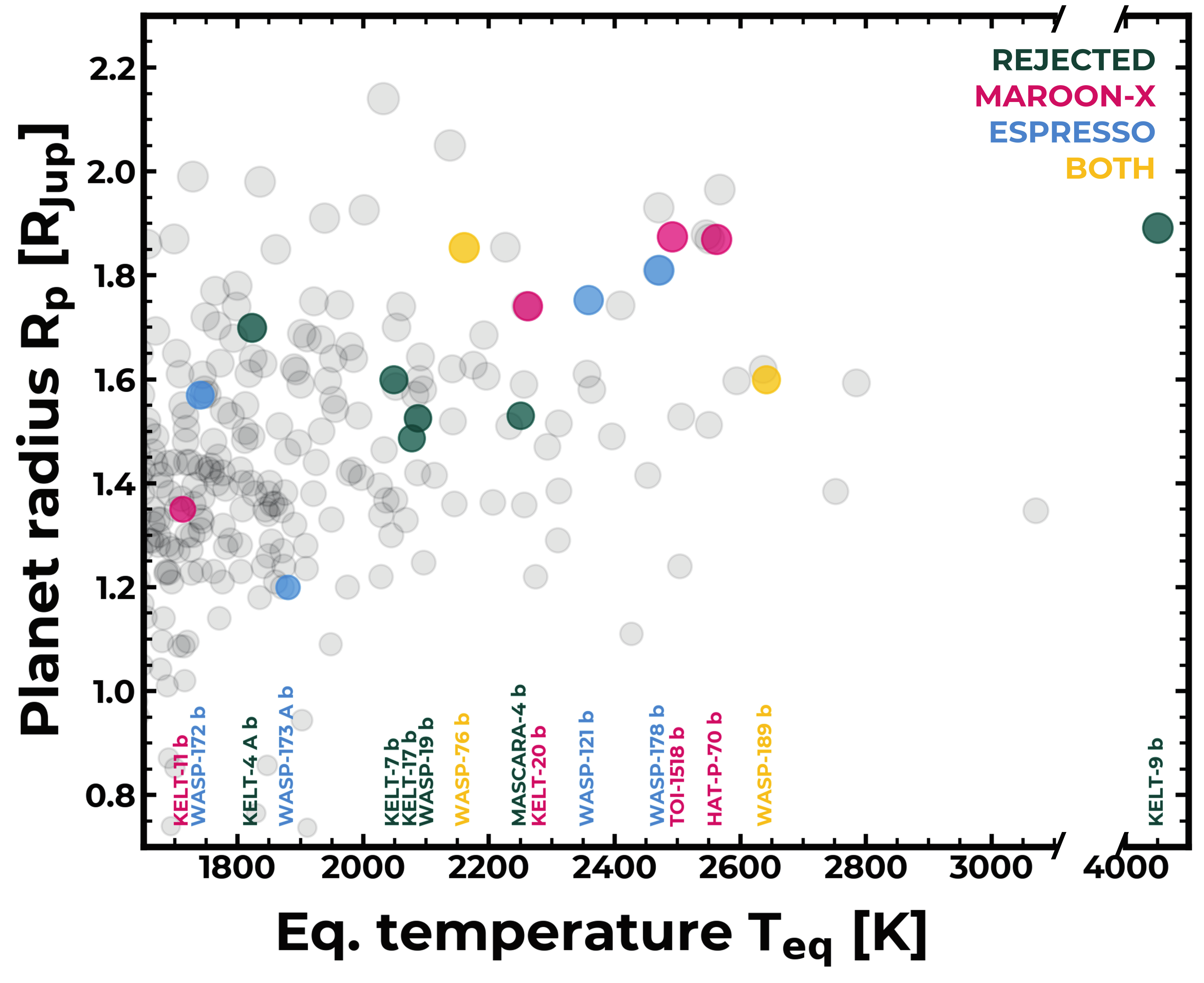}
    \caption{Overview of the exoplanet population in the gas-giant regime with radii between 0.75 and 2.5 R$_{\rm Jup}$, extracted from the NASA Exoplanet Archive on 12 June 2026. The parameter space shown corresponds to hot to ultra-hot Jupiters with equilibrium temperatures between 1500 and 4100~K. Planets included in our sample are colour-coded by the instrument with which they have been observed, in red (MAROON-X), blue (ESPRESSO), and yellow (observed with both). Rejected targets are displayed in green. We excluded MASCARA-4~b, KELT-4~A~b, and KELT-7~b due to orbiting around pulsating stars, WASP-19~b because of low-S/N observations from its relatively faint host star, KELT-17~b because of its chemically peculiar Am host star, and KELT-9~b due to probing a different chemical regime given its high temperature.}
    \label{fig:population}
\end{figure}

Figure~\ref{fig:population} illustrates the location of our targets in the radius–temperature plane relative to the broader gas-giant population. A full observational log of included and excluded systems is provided in the Appendix, see Table~\ref{tab:observation_log}. The data were obtained either directly from principal investigators of the respective observing programmes or retrieved from the ESO/Gemini archives if previously used in a peer-reviewed publication.

ESPRESSO observations were reduced with the dedicated pipeline (v3.2.0) within the \texttt{esoreflex} environment (v2.11.5) provided by ESO and the ESPRESSO consortium. For the cross-correlation analysis, we used the non-blaze-corrected, two-dimensional order-by-order spectra from fibre A (S2D products). However, for the telluric correction, the flux-calibrated one-dimensional spectra from fibre A (S1D products) provide a better continuum fit and were used instead and then interpolated on the S2D orders. The ESPRESSO pipeline applies Barycentric Earth Radial Velocity (BERV) corrections directly to the S2D spectra.  

MAROON-X data were reduced using the standard pipeline described in \citet{seifahrt_-sky_2020}, which produces order-by-order spectra analogous to ESPRESSO S2D products. Due to differences in readout times between the red and blue arms, exposures are synchronised at the start but recorded separately, and were therefore treated as independent observations. Since the pipeline does not produce flux-calibrated one-dimensional spectra, we stitched the reduced data to construct a one-dimensional time series for telluric correction \citep{prinoth_time-resolved_2023}. Unlike ESPRESSO, BERV corrections are not applied during reduction, and were instead applied during the cross-correlation analysis.

\section{Methods}
We carried out a homogeneous reanalysis of all datasets using the open-source package \texttt{tayph} \citep[e.g.][]{hoeijmakers_atomic_2018,hoeijmakers_hot_2020,prinoth_titanium_2022,borsato_mantis_2023}. It provides routines for telluric correction, masking, outlier rejection, and cross-correlation of the form
\begin{equation}
    C(v,t) = \sum_{i = 0}^{N} F_{i}(t)T_i(v),
    \label{eq:cc}
\end{equation}
where $T(v)$ represents the cross-correlation template shifted to a radial velocity $v$ and is normalised such that $\sum_{i=0}^{N} T_i(v) = 1$ with $i$ denoting the spectral pixels,  $F_{i}(t)$ is the observed flux at time $t$. We opted to perform the Rossiter-McLaughlin and systemic velocity correction outside of the \texttt{tayph}-framework, whereas the former is normally part of it. 

\subsection{Telluric correction and masking}

We corrected for contamination by Earth's atmosphere with the standalone version of \texttt{molecfit} \citep[v1.5.9;][]{smette_molecfit_2015,kausch_molecfit_2015}, fitting each exposure individually in spectral regions dominated by water vapour and \ch{O2} absorption. The resulting models were interpolated onto the same wavelength grid\footnote{Note that this included velocity corrections to account for the BERV.} as the order-by-order spectra (S2D products) and divided out. We then shifted to the stellar rest frame, where the spectral lines of the host star are stationary. This shift accounts for the stellar reflex motion due to the orbiting planet but excludes the systemic velocity at this point, given that it only introduces a constant offset. 
To mitigate residual features, we applied a sigma-clipping algorithm following \citet{hoeijmakers_hot_2020}. With the spectra shifted into the stellar rest frame, we construct transmission spectra by dividing each exposure by the out-of-transit mean stellar spectrum. We then compute the median absolute deviation within 40-pixel windows in the resulting residual (transmission) spectra, and flag all $5\sigma$ outliers. The resulting mask is subsequently applied to the non-normalised spectra used in the cross-correlation analysis. This ensures that outlier rejection is based on the residual spectra, while preserving the noise properties and airmass dependencies of the original flux spectra during the cross-correlation step. Additionally, we mask pixels where the telluric contamination reduces the flux by more than 50\% to avoid excessive noise on the cross-correlation signal. 
{\tt molecfit} successfully corrects the telluric water lines around the sodium doublet (589–589.6~nm) to a level well below the pre-correction noise, with residuals typically at or below a few percent of the noise level.

\subsection{Cross-correlation analysis}
\label{sec:cc}

\begin{figure}[t]
    \centering
    \includegraphics[width=\linewidth]{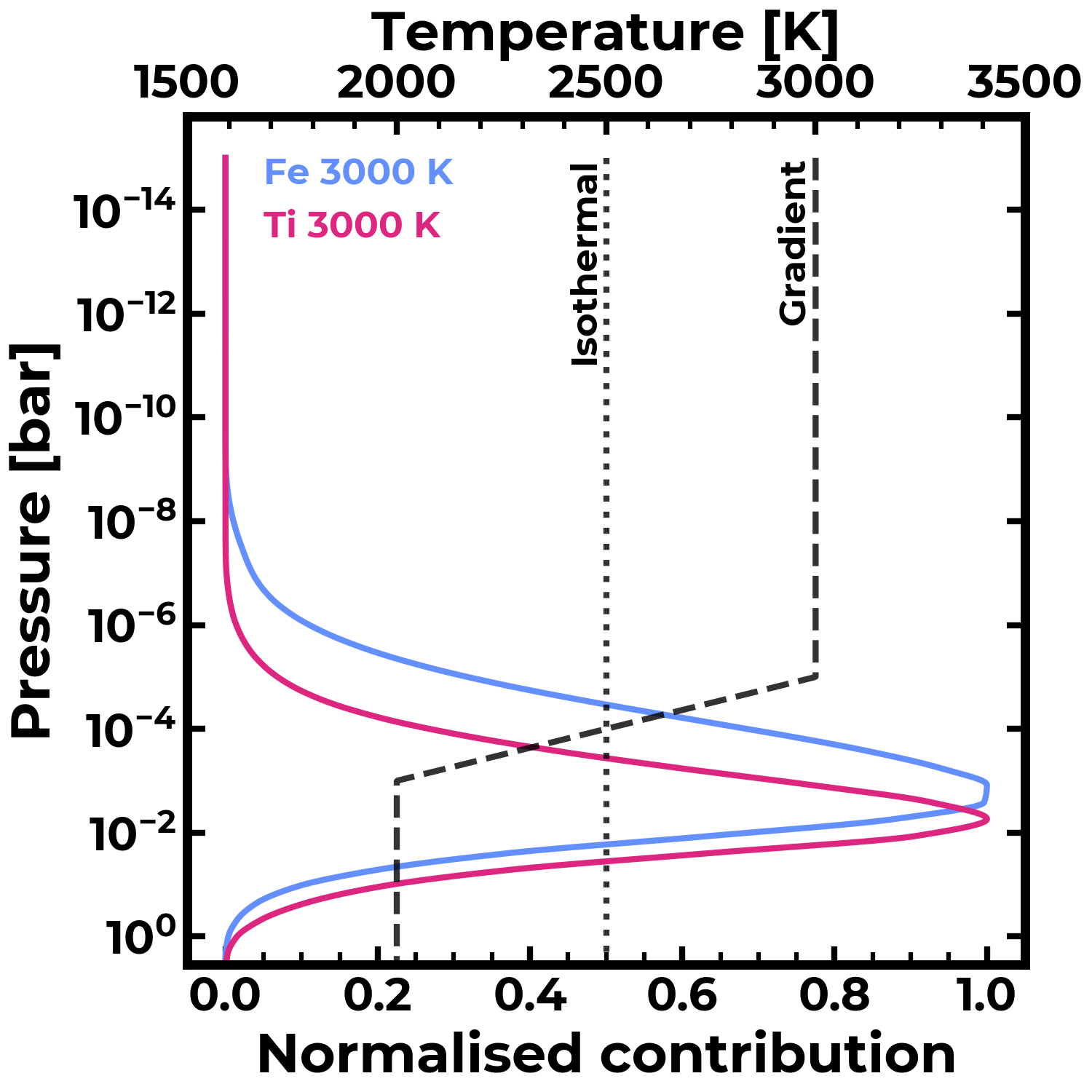}
    \caption{Contribution functions of the cross-correlation templates and considered temperature-pressure profiles for the forward models. The Fe template probes slightly higher compared to the Ti template. 
    }
    \label{fig:cfs}
\end{figure}

We performed the cross-correlation analysis for all planets in our sample with templates for \ion{Fe}{} and \ion{Ti}{} calculated at \SI{3000}{\kelvin} from \citet{kitzmann_mantis_2023}. These templates assume an isothermal atmosphere at \SI{3000}{\kelvin} in hydrostatic and chemical equilibrium. The templates were calculated for a 'standard' inflated ultra-hot Jupiter with R$_p$ = 1.5~R$_{\rm Jup}$, and a base pressure at the bottom of the atmosphere of 10~bar. To achieve a homogeneous population study, we used the same template across all data sets even if this may result in lower detection significances for individual planets. These templates probe roughly similar pressure levels in the atmosphere, see Fig.~\ref{fig:cfs}. After the cross-correlation with the template over a radial velocity range from \num{-1000} to \SI{1000}{\km\per\second} in steps of \SI{1}{\km\per\second}, we divided the mean out-of-transit cross-correlation function to remove the signal from the host star. Lastly, we applied a high-pass filter with a FWHM of \SI{100}{\km\per\second} to remove any residual broad-band variation. 

\subsection{Rossiter–McLaughlin correction}
When a planet transits in front of its host star, it consecutively covers parts of the rotating stellar disc.  This obscured area introduces distortions in the stellar lines, called the Rossiter–McLaughlin effect \citep[RM][]{rossiter_detection_1924,mclaughlin_results_1924}.
We corrected for the RM effect following the method described by \citet{prinoth_titanium_2025}, using the \texttt{StarRotator} code \citep[see also][and Hoeijmakers, in prep.]{prinoth_time-resolved_2023,jens_hoeijmakers_2024_13789136,lam_secrets_2024}. This approach models the average stellar cross-correlation function as a Gaussian centred on the systemic velocity. The Gaussian fit is shifted by the component of the stellar rotational velocity projected along the line of sight, which corresponds to the region of the stellar surface occulted by the transiting planet.
The local velocity of each stellar surface element depends on the projected stellar rotational velocity, $v\sin{i}$, and the planet’s position on the stellar disc, which is governed by the projected orbital obliquity, $\lambda$. For a full derivation of the velocity field and model geometry, we refer to \citet{prinoth_atlas_2024}. The planet-to-star radius ratio was held fixed due to its degeneracy with the amplitude and width of the stellar cross-correlation function (the Gaussian).
We performed the parameter inference within a Bayesian framework using \texttt{pymultinest} \citep{buchner_x-ray_2014}. For details of the implementation and a representative application, see \texttt{StarRotator} \citep{jens_hoeijmakers_2024_13789136} and the full methodology in \citet{prinoth_titanium_2025}.

Following the RM correction, we visually inspected the corrected cross-correlation maps for each dataset and template combination to assess the quality of the removal. In general, the RM signal was removed down to the noise level. However, in a small number of cases, individual exposures displayed significant residual structures after correction, typically associated with anomalous flux variations or poorly corrected systematics. Such exposures were excluded from the subsequent analysis to avoid introducing spurious features into the resulting cross-correlation functions. In total, no more than three exposures were excluded for any individual dataset. These exclusions occurred only for the Fe analysis, where the RM signal is strongest, while for Ti the RM contribution is weak to negligible.

\subsection{Velocity correction}
We measured the systemic velocity for each transit observation by fitting a rotationally broadened stellar line \citep{gray_observation_2008} to the average cross-correlation function of all out-of-transit exposures. To obtain the cross-correlation signal of the star, we used a suitable PHOENIX atmosphere model \citep{husser_new_2013} based on the stellar parameters for the effective temperature $T_{\rm eff}$ and surface gravity $\log{g}$ in Table~\ref{tab:params} as a template, broadened to the line-spread function of the instrument.  The velocity shift measured for each transit was applied to align all exposures of a given time series into the true rest frame of the system.
Fitting the systemic velocity for each individual time series assures that the signals are truly aligned and not caused by relative wavelength offsets between spectrographs and dates. However, since we do not study any dynamical aspects of the sample, this velocity correction has a minor impact on the fit, but is performed here for consistency. A detailed description of the systemic velocity fit can be found in \citep{seidel_magnetic_2026}. Following the shift of the cross-correlation signal by the constant offset of the system velocity, we produce the \kpvsys maps by assuming different values for the orbital velocity $K_p$  ranging from 0 to \SI{400}{\km\per\second}. For each of these steps (\SI{1}{\km\per\second}), we shifted the spectra to this assumed planetary rest frame 

\begin{equation}
    v_{\rm planet} = K_p\sin(2 \pi \phi) \sin{i} + v_{\rm sys} - v_{\rm BERV},
    \label{eq:planet_v}
\end{equation}

where $\phi$ is the orbital phase ($\phi = 0$ is the transit centre), $i$ is the orbital inclination, $v_{\rm sys}$ is the systemic velocity for the given observation, and $v_{\rm BERV}$ is the barycentre velocity. Note that we corrected for $v_{\rm BERV}$ and $v_{\rm sys}$ before, so they are only mentioned in Eq.~\eqref{eq:planet_v} to showcase the rest frame of the planet with respect to the observer's rest frame.
Because all these planets orbit on extremely short orbits (P $\leq$ \SI{6}{days}), their orbits should have had enough time to be circularised over the life time of the systems caused by strong tidal interactions \citep[e.g.][]{1981Hut,Matsumura_2010}. Hence, it is also assumed that all planets orbit on circular orbit, and no eccentricity is included in Eq.~\eqref{eq:planet_v}.

\subsection{Model prediction and comparison}
\label{sec:models}

For each planet with a Fe and/or Ti detection ($\geq\,5\sigma$), we generated a suite of forward-model transmission spectra using \texttt{petitRADTRANS} v2 \citep{molliere_petitradtrans_2019}, based on the planetary and stellar parameters listed in Table~\ref{tab:params}. Atomic line opacities for Fe and Ti were taken from the VALD3 database \citep{ryabchikova_major_2015}, and continuum opacity sources included collision-induced absorption from \ch{H2}-\ch{H2} \citep{Borysow_2001,Borysow_2002} and \ch{H2}-\ch{He} \citep{Borysow_1988}, bound-free and free-free absorption from \ch{H-} \citep{gray_observation_2008}, and Rayleigh scattering by \ch{H2} \citep{Dalgarno_1962} and \ch{He} \citep{Chan_1965}. 

We assumed the atmospheres to be one-dimensional and in hydrostatic and chemical equilibrium. Two temperature–pressure prescriptions were considered to capture plausible thermal structures probed via transmission spectroscopy. In the first scenario (“isothermal”), the atmosphere is held at a constant temperature $T$, varied from \num{2000} to \SI{4000}{\kelvin} in \SI{200}{\kelvin} steps. In the second scenario (“gradient”), a simplified inverted profile was adopted in which temperature varies linearly with log-pressure between $10^{-5}$ and $10^{-3}$~bar \citep[as has been adopted on the daysides of ultra-hot Jupiters, e.g.][see their Fig.~7]{Bonidie_2026}, transitioning from $T+500$~K at low pressures to $T-500$~K at higher pressures, and remaining isothermal outside this range. These scenarios bracket a range of plausible upper-atmosphere temperature structures. Example T–p profiles for $T = \SI{2500}{\kelvin}$ are shown in Fig.~\ref{fig:cfs}. 

Based on these T–p profiles, we computed the equilibrium volume mixing ratios using \texttt{FastChem Cond} \citep{stock_fastchem_2018,stock_fastchem_2022,Kitzmann_2023}, including condensation via the rainout prescription from \citet{marley_clouds_2013}. To explore the impact of elemental abundance variations, Fe and Ti abundances were scaled by factors 1 (nominal), $10^{-1}$, and $10^{-2}$  relative to stellar metallicity (see Table~\ref{tab:params}) prior to equilibrium calculations. 

We injected these forward models into the observed time-series data at the expected projected orbital velocity $K_p = \frac{2\pi a}{P} \sin{i}$ and systemic velocity $v_{\rm sys}$ following the methodology described in \citet{hoeijmakers_high-resolution_2020} and subsequent studies \citep[e.g.,][]{prinoth_time-resolved_2023,hoeijmakers_mantis_2024}. We then performed the same cross-correlation analysis as in section~\ref{sec:cc} on the dataset including the injected forward models, yielding the recovered cross-correlation amplitudes expected for each temperature and abundance scenario under the same observational conditions as the actual data was taken. 

\subsection{Cross-correlation metric \metric}
\label{sec:metric}

To quantify the relative trend of Ti and Fe while minimising system-dependent effects, we define a logarithmic depletion metric based on the recovered cross-correlation amplitudes. Absolute signal strengths depend strongly on factors such as planetary radius, surface gravity, atmospheric scale height, stellar brightness, signal-to-noise ratio, and instrumental systematics, making direct comparisons across different planets or instruments unreliable.

By taking the ratio of Ti to Fe signals within the same dataset, many of these shared dependencies largely cancel out. To place all planets on a common scale, this ratio is further normalised by a fixed reference value derived from isothermal model spectra at \SI{2600}{\kelvin}. The metric \metric is thus defined as:

\begin{equation}
    \Delta_{\text{Ti} - \text{Fe}} =
    \log_{10}\left(\frac{S_{\rm Ti, obs}}{S_{\rm Fe, obs}}\right)
    -
    \log_{10}\left(\frac{S_{\rm Ti, ref\ at\ 2600\,K}}{S_{\rm Fe, ref\ at\ 2600\,K}}\right),
    \label{eq:metric}
\end{equation}

where $S_{\rm X, obs}$ denotes the observed cross-correlation peak amplitude for species $X$ ($X \in$ [Fe, Ti]).

The \SI{2600}{\kelvin}reference does not represent the atmospheric temperature of any individual planet, but instead defines a fixed baseline used consistently throughout the analysis. Importantly, the same reference is adopted for all temperature--pressure prescriptions considered in this work, including both isothermal and temperature-gradient models. In practice, all forward-model spectra, irrespective of their temperature structure, are projected onto the same fixed reference template at \SI{2600}{\kelvin} when computing \metric. This ensures that all measurements and model predictions are evaluated relative to the same baseline and that differences in atmospheric structure affect only the recovered signal amplitudes themselves, rather than the definition of the normalisation.

The choice of \SI{2600}{\kelvin} is motivated by the fact that Ti and Fe are expected to exhibit broadly similar behaviour in this regime, with minimal contributions from ionisation effects and TiO formation at the pressures probed (see Fig.~\ref{fig:cfs}), thereby providing a stable anchoring point for the comparison. The metric should therefore be interpreted as a relative deviation with respect to a single fiducial model reference point, rather than as a per-planet equilibrium chemistry residual. While injected models are planet-specific, all cross-correlations are computed using the same templates (see Section~\ref{sec:cc}), ensuring a homogeneous basis for the population-level analysis.

To compare the observed trends with model predictions, we evaluate the same metric on the forward models by replacing $S_{\rm X, obs}$ with the corresponding cross-correlation strengths recovered from the injected signals. This yields predicted values of \metric for both isothermal and temperature-gradient models across a range of atmospheric temperatures (\num{2000}--\SI{4000}{\kelvin}) and elemental abundance scalings (1, $10^{-1}$, $10^{-2}$) for each planet.

\section{Results and Discussion}
\label{sec:resultsdiscussion}

\subsection{Detections and population-level trend of Fe and Ti}

We performed a homogeneous high-resolution cross-correlation analysis of neutral iron and titanium across a sample of ten hot and ultra-hot Jupiters. We detect Fe at high significance ($>5\sigma$) in seven planets (\fedetections), while Ti is detected in only four (\tidetections; see Fig.~\ref{fig:fe_det} and Fig.~\ref{fig:ti_det}). Iron therefore appears ubiquitous in all planets in our sample with $T_{\rm eq}$ $\gtrsim$ 2200\,K, whereas titanium is only detected in a subset of the hottest ($T_{\rm eq}$ $\gtrsim$ 2350\,K) planets. Although all reported species have previously been detected in the literature \citep[e.g.,][]{ehrenreich_nightside_2020,stangret_detection_2020,BelloArufe_mining_2022,prinoth_titanium_2022,damasceno_atmospheric_2024-1,simonnin_time_2024,prinoth_titanium_2025,Sun2026}, our uniform analysis enables a direct and unbiased comparison of their relative trend across the population.

No significant detections are found for KELT-11~b, WASP-172~b, and WASP-173~A~b. The comparatively low equilibrium temperature of KELT-11~b ($T_\mathrm{eq} \sim \SI{1700}{\kelvin}$) likely suppresses neutral atomic species, while the higher surface gravity of WASP-173~A~b reduces its atmospheric scale height and thus the strength of transmission signatures. WASP-172~b does show a tentative Fe signal \citep{Seidel_Prinoth_2023}, although it remains below the detection threshold in our re-analysis. These cooler planets ($T_\mathrm{eq} \lesssim \SI{2000}{\kelvin}$) probe the regime near the expected condensation boundaries of refractory elements and therefore provide important constraints on their removal from the gas phase. 
Our non-detections of Ti on WASP-178~b is likely due to the lower data quality and overlap of the RM with the planetary trace in these data, as discussed in \citet{damasceno_atmospheric_2024-1} and evidenced by the weaker observed Fe signal (Figure~\ref{fig:fe_det}).  Indeed, given its equilibrium temperature between that WASP-121~b and TOI-1518~b, both with Ti detections, we expect Ti to also be present in the gas phase on WASP-178~b. However, more observations will be needed to confirm this. 

\begin{figure}
    \centering
    \includegraphics[width=\linewidth]{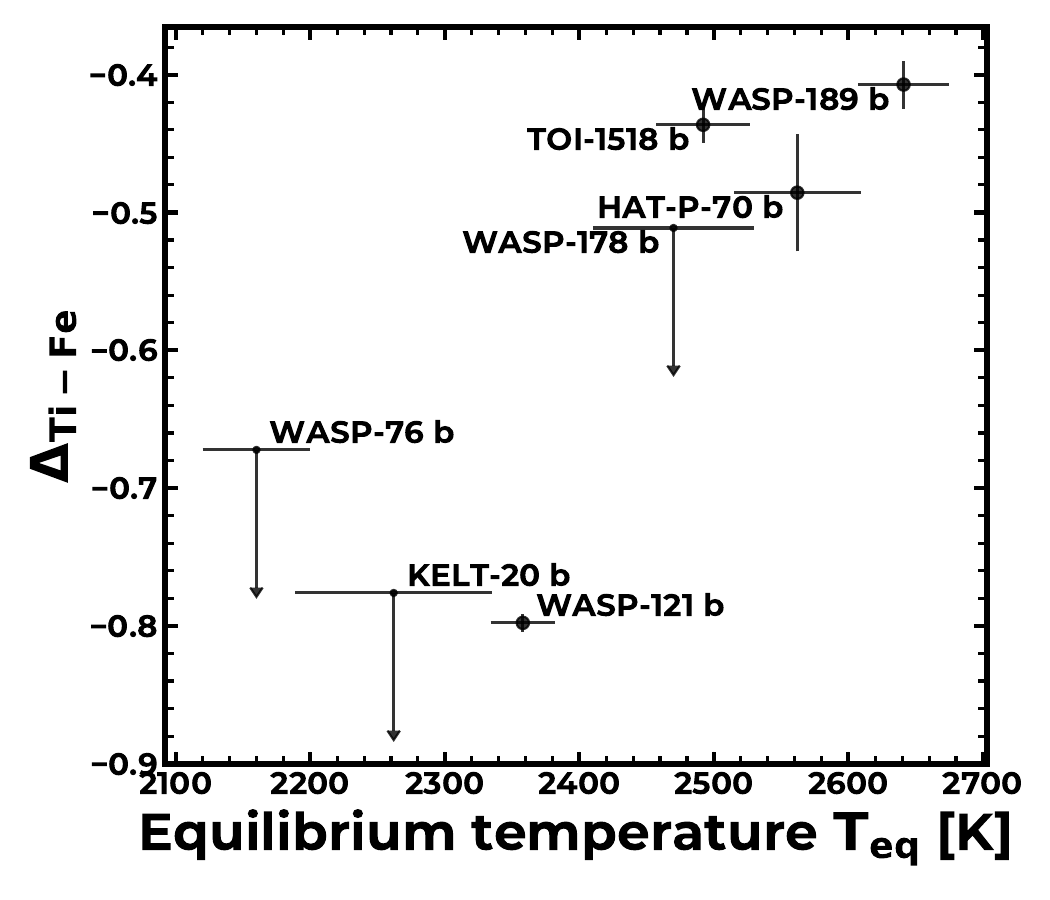}
    \caption{Observed \metric across the sub-sample of the population with Fe and/or Ti detections. The \metric shows a clear downward trend towards colder equilibrium temperatures, indicating that the degree to which Ti is weaker than Fe compared to model prediction is more severe on colder planets. The uncertainties of the observed \metric are computed via Gaussian error propagation of the fitted peak of the cross-correlation map in Eq.~\eqref{eq:metric}.
    }
    \label{fig:metric_womodels}
\end{figure}

To quantify the the trend across the sample, we use the cross-correlation metric \metric (Eq.~\ref{eq:metric}), which measures the Ti-to-Fe signal ratio relative to a reference model and largely removes system-dependent observational effects. The \metric measured across our sample exhibits a clear decrease toward lower equilibrium temperatures (Fig.~\ref{fig:metric_womodels}). While Fe remains detectable across most of the temperature range, Ti absorption declines rapidly below $T_\mathrm{eq} \sim \num{2200}-\SI{2400}{\kelvin}$.

This population-level trend demonstrates that titanium is systematically removed from the observable atomic phase at higher temperatures than iron. The trend is consistent with previous retrieval studies \citep[e.g.,][]{gandhi_retrieval_2023, maguire_high-resolution_2023, pelletier_vanadium_2023, Pelletier_2026, prinoth_titanium_2025}, which report Ti abundances depleted by $\log(X/X_\star) \sim -1$ or more, while Fe remains close to stellar values.

\begin{figure*}[t]
    \centering
    \includegraphics[width=\linewidth]{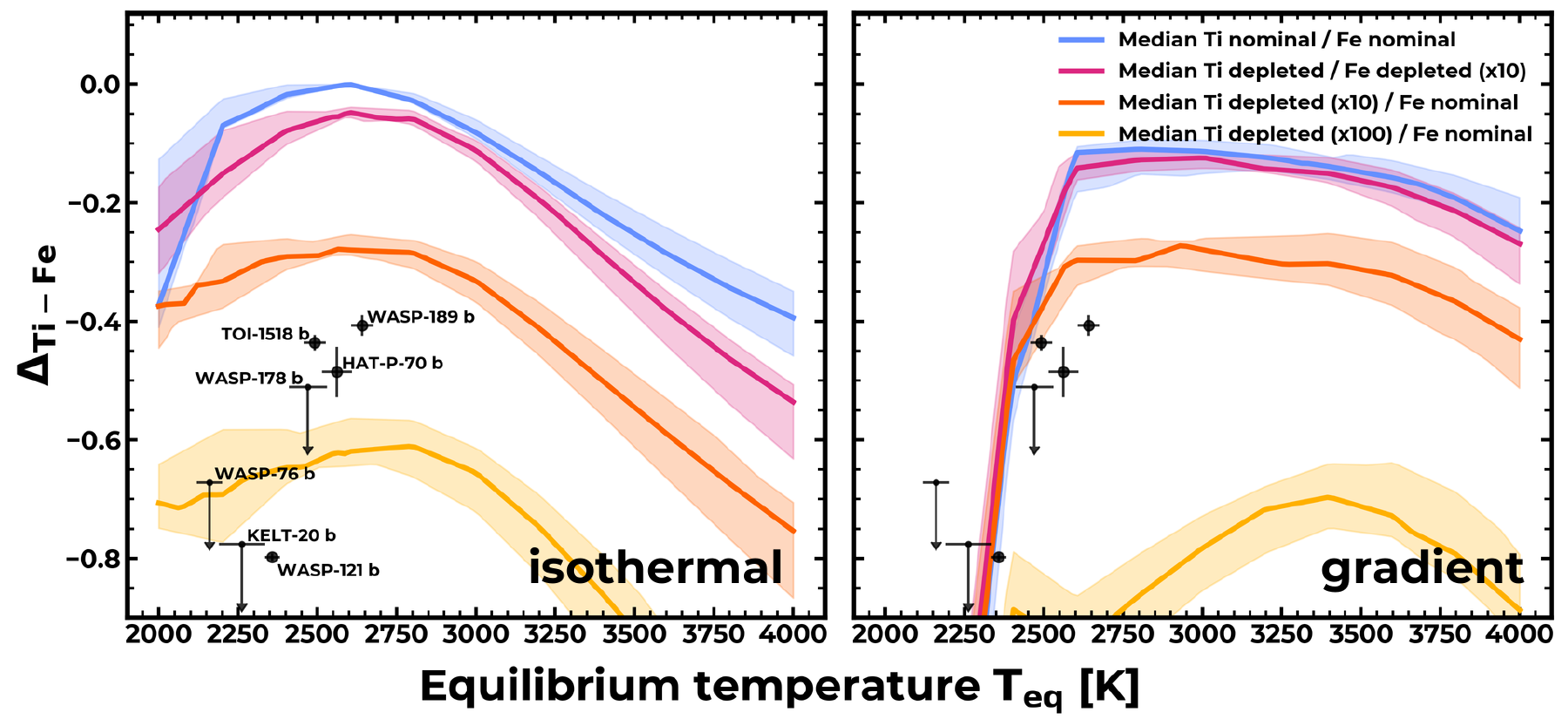}
    \caption{Model comparison for isothermal (left) and gradient (right) models. The upper limits correspond to 3$\sigma$ limits based on the noise in the cross-correlation map. The model medians are computed from evaluating the models in Eq.~\eqref{eq:metric} for all planets. The shaded regions indicate the 84th percentiles, showing that this metric effectively cancels out any major planet-specific dependencies, meaning that the overall trend of the metric is captured by all models together. 
    } 
    \label{fig:metric}
\end{figure*}

\subsection{Constraints from forward models}

To interpret this observed population trend, we compare the measured values of \metric to predictions from a suite of forward models (see Section~\ref{sec:models}). Figure~\ref{fig:metric} shows the model predictions averaged over all planets at each temperature. The small scatter between individual planets (captured in the 84th percentile region) demonstrates that \metric effectively removes most system-dependent effects, enabling a direct comparison between models and observations.

The isothermal models predict only a weak dependence of \metric on temperature across the parameter space probed by our sample. At higher temperatures, \metric decreases due to the onset of ionisation beyond \SI{2600}{\kelvin}, which preferentially reduces the neutral Ti signal because of its lower ionisation energy relative to Fe \citep[e.g.][]{prinoth_time-resolved_2023}. At lower temperatures, however, \metric remains nearly constant. Although Ti is partially converted into TiO, this transition occurs gradually over a broad pressure range, leaving a substantial fraction of atomic Ti distributed across the atmosphere (see Fig.~\ref{fig:partitioning}). As a result, the observable Ti signal is only weakly affected, and isothermal atmospheres cannot reproduce the strong decline in Ti relative to Fe observed across the population under the assumption of uniform Ti depletion.

Thus reproducing the observed trend with isothermal models therefore requires invoking a temperature-dependent depletion of titanium, with cooler planets exhibiting progressively stronger depletion relative to Fe. This scenario naturally points toward cold-trapping processes that become more efficient at lower equilibrium temperatures, for example through enhanced condensation of Ti-bearing species on the nightside \citep{parmentier_3d_2013}. In this picture, the observed population trend directly reflects a temperature-dependent removal of Ti from the atmosphere.

In contrast, models including a temperature gradient reproduce the observed trend without requiring a temperature-dependent depletion. These models produce a rapid decline in \metric over a narrow temperature range and therefore provide a much better match to the data. In such atmospheres, temperatures decrease with increasing pressure over the region probed by the cross-correlation templates. Because the Ti template probes slightly deeper atmospheric layers than Fe (see Fig.~\ref{fig:cfs}), it becomes more sensitive to cooler regions of the atmosphere. In these deeper layers, Ti is efficiently converted into TiO and subsequently removed from the gas phase through condensation (e.g. into Ti-bearing condensates such as \ch{CaTiO3}). This process both depletes the available Ti budget more significantly compared to Fe (see Fig.~\ref{fig:partitioning}) and confines the remaining atomic Ti to a narrow region. This differential response naturally produces the observed decline in \metric.

At lower temperatures, the gradient models become increasingly insensitive to the overall Ti abundance. In this regime, most Ti is already efficiently recombined into TiO (or rained out completely), and the residual atomic Ti signal is primarily set by the local chemical balance rather than the total Ti inventory, making it difficult to distinguish between nominal and depleted abundances. In contrast, the hottest planets in our sample suggest that Ti is globally depleted relative to Fe by roughly an order of magnitude. This finding disfavours a purely temperature-dependent depletion scenario and instead points toward an overall depletion of Ti relative to Fe across the population.

\subsection{Breaking degeneracies and future prospects}

Distinguishing between intrinsic abundance variations and atmospheric processes such as cold trapping requires additional observational constraints.

One complementary avenue is the detection of TiO in planets with equilibrium temperatures of $\sim$\num{2000}--\SI{2500}{\kelvin}, where TiO formation is expected to be efficient. In this regime, an anti-correlation between atomic Ti and molecular TiO would provide a direct signature of chemical partitioning. However, current limitations in TiO line lists may complicate or even prevent such measurements, except for the most favourable planets \citep{hoeijmakers_search_2015, mckemmish_exomol_2019, prinoth_titanium_2022}. Ground-based observations are particularly affected by inaccuracies in these line lists, whereas space-based observations are less sensitive to such issues because they do not resolve the dense forest of individual spectral lines, which makes them prone to being confused with other species or even reflected light \citep{Pelletier_2026}.

Extending the sample to higher equilibrium temperatures provides a complementary test. If the observed trend is primarily driven by TiO formation and condensation associated with temperature gradients, \metric should flatten at sufficiently high temperatures, where titanium remains predominantly atomic (as suggested by Fig.~\ref{fig:metric}). At the same time, however, ionisation becomes significant and may further reduce the observable neutral Ti abundance. Nonetheless, exploring higher temperatures should help confirm or rule out (partial/temperature-dependent) cold trapping as the underlying cause. Testing this region of parameter space would also clarify whether these planets follow the predicted overall depletion in Ti/Fe.

Finally, high-resolution emission spectroscopy probes deeper and hotter atmospheric layers than transmission spectroscopy and is more sensitive to the temperature–pressure structure. Joint transmission and emission measurements of the same planets would provide a powerful means of disentangling vertical structure effects from global abundance variations. Together, these observations will be essential for establishing a comprehensive picture of refractory element chemistry in strongly irradiated exoplanet atmospheres.

\section{Conclusions}
\label{sec:conclusions}

We have presented a homogeneous high-resolution spectroscopic analysis of iron and titanium absorption in ten hot and ultra-hot Jupiter atmospheres observed with ESPRESSO and MAROON-X. Using a consistent methodology and the relative cross-correlation metric \metric, we directly compare the observed trend of these species across a critical temperature range for refractory chemistry. We identify a clear population-level trend in which \metric declines rapidly toward lower equilibrium temperatures. 

Forward-model comparisons indicate that reproducing this trend requires either of two scenarios. First, assuming isothermal temperature-pressure profiles, Ti is progressively depleted relative to Fe in cooler planets, consistent with selective cold-trapping processes. In this case, the hottest planets show little, while the coldest planets exhibit the strongest Ti depletion. Second, invoking non-isothermal profiles with atmospheric temperature gradients leads to efficient conversion of Ti into TiO in deeper layers, followed by condensation into Ti-bearing species, which removes Ti from the observable atmosphere. While this scenario naturally reproduces the observed trend, the hottest planets still indicate an additional overall Ti depletion of roughly a factor of 10 relative to Fe that cannot be explained by this mechanism alone. These findings suggest that molecular partitioning, ionisation, condensation, and cold-trapping processes collectively shape the observable refractory chemistry of ultra-hot Jupiter atmospheres.

Although current observations cannot fully distinguish between these scenarios, expanding homogeneous high-resolution surveys to larger and more diverse samples will be crucial. This includes planets orbiting fainter host stars, which will become accessible with high-resolution spectrographs on the Extremely Large Telescope or using ESPRESSO's 4UT mode \citep[e.g.][]{borsa_atmospheric_2021,seidel_vertical_2025,prinoth_titanium_2025,Vaughan_2026}, or more challenging host stars \citep[e.g., active stars or pulsators]{Jiang2026}. Such efforts will be key to breaking these degeneracies and achieving a comprehensive understanding of refractory element chemistry in strongly irradiated giant planets.

\begin{acknowledgements}
    We thank the anonymous referee for their comments and suggestions, which have improved the clarity of the manuscript. The authors acknowledge the ESPRESSO project team for its effort and dedication in building the ESPRESSO instrument, as well as the project team and support staff of MAROON-X. This work relied on observations collected at the European Organisation for Astronomical Research in the Southern Hemisphere and at the international Gemini Observatory, a program of NSF NOIRLab, which is managed by the Association of Universities for Research in Astronomy (AURA) under a cooperative agreement with the U.S. National Science Foundation, on behalf of the Gemini partnership of Argentina, Brazil, Canada, Chile, the Republic of Korea, and the United States of America. This work was enabled by observations made from the Gemini North telescope, located within the Mauna Kea Science Reserve and adjacent to the summit of Mauna Kea. We are grateful for the privilege of observing the Universe from a place that is unique in both its astronomical quality and its cultural significance.
    B.P. acknowledges partial financial support from The Fund of the Walter Gyllenberg Foundation of the Royal Physiographic Society of Lund for funding the collaboration stay in Nice, France, to enable this project. V.P. acknowledges financial support from the French National Research Agency (ANR) project EXOWINDS (ANR-23-CE31-0001-01). This project has been carried out within the framework of the National Centre of Competence in Research PlanetS supported by the Swiss National Science Foundation under grant 51NF40\_205606. S.P.\ acknowledges the financial support of the SNSF.  F.D. acknowledges funding from the French National Research Agency (ANR) project ExoATMO (ANR-25-CE49-6598).
\end{acknowledgements}

\bibliographystyle{aa}
\bibliography{references}

\appendix
\onecolumn

\section{Detection maps}

\begin{figure*}[h!]
    \centering
    \includegraphics[width=\linewidth]{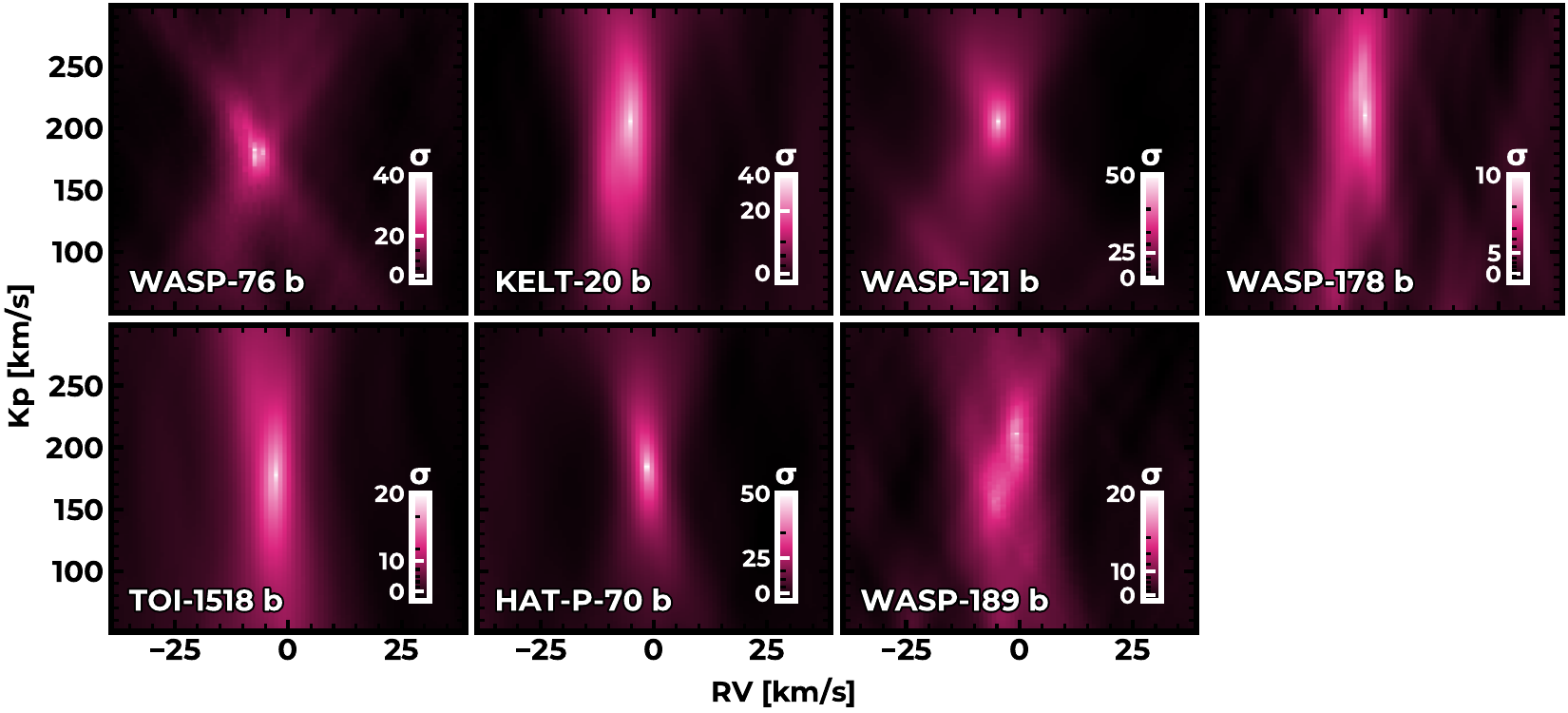}
    \caption{\kpvsys diagrams showing detected Fe absorption in the sample. Each panel displays the stacked \kpvsys diagram after correcting individual observations for their system velocities. The colour bars indicate the detection significance, calculated by dividing the diagram by its standard deviation outside the velocity ranges affected by the star, planet, Rossiter–McLaughlin effect, and tellurics.}
    \label{fig:fe_det}
\end{figure*}

\begin{figure*}[h!]
    \centering
    \includegraphics[width=\linewidth]{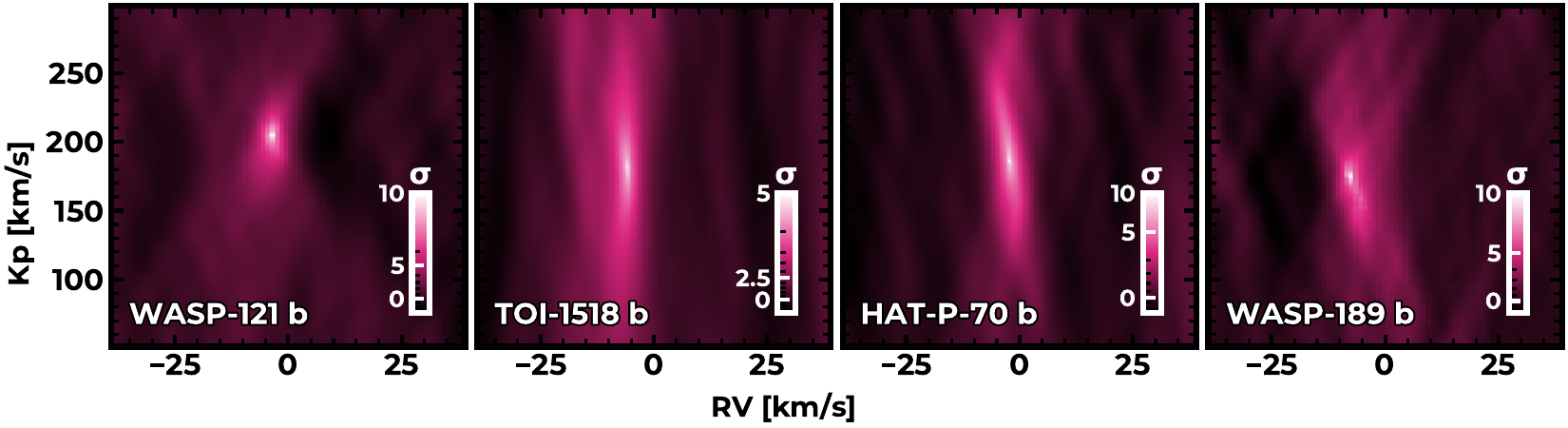}
    \caption{Same as Fig.~\ref{fig:fe_det} but for Ti.}
    \label{fig:ti_det}
\end{figure*}

\newpage

\section{Stellar and planetary parameters for forward models}

\begin{table}[h!]
    \renewcommand{\arraystretch}{1.5}
    \caption{Overview of the stellar and planetary parameters adopted in this study.}
    \centering
    {\footnotesize
    \begin{tabular}{p{0.14\textwidth}p{0.1\textwidth}p{0.1\textwidth}p{0.1\textwidth}p{0.1\textwidth}p{0.1\textwidth}p{0.1\textwidth}p{0.1\textwidth}}
    & WASP-76 b & KELT-20 b & WASP-121 b & WASP-178 b & TOI-1518 b & HAT-P-70 b & WASP-189 b \\
    \midrule
    \multicolumn{3}{l}{Stellar parameters} &&&& \\
    \midrule
    $T_{\rm eff}$ [K] &  $6329 \pm 65$  & \numpm{8720}{+250}{-260} & $6459 \pm 140$ & $9360 \pm 150$ & \numpm{7299}{+98}{-100} & \numpm{8450}{+540}{-690} & $8000 \pm 80$ \\
    log$_{\rm g}$ [cgs] & $4.196 \pm 0.106$ & \numpm{4.290}{+0.017}{-0.020}& \numpm{4.242}{+0.011}{-0.012} & $4.31 \pm 0.04$ & $4.136 \pm 0.011$ & \numpm{4.181}{+0.055}{-0.063} & $3.9 \pm 0.2$ \\
    $R_{\rm s}$ [$R_\odot$] & $1.756 \pm 0.071$ & \numpm{1.565}{+0.057}{-0.064} & $1.458 \pm 0.030$ & $1.67 \pm 0.07$ & $1.942 \pm 0.043$ & \numpm{1.858}{+0.119}{-0.091} & $2.36 \pm 0.03$ \\
    Metallicity [Fe/H] & $0.366 \pm 0.053$ & \numpm{-0.29}{+0.22}{-0.36} & $0.13 \pm 0.09$ & $0.21 \pm 0.16$ & $-0.10 \pm 0.12$ & \numpm{-0.0590}{+0.0750}{-0.0880} & $0.29 \pm 0.13$ \\
    \midrule
    \multicolumn{3}{l}{Planetary parameters} &&&& \\
    \midrule
    $R_{\rm p}$ [$R_{\rm Jup}$] & \numpm{1.854}{+0.077}{-0.076} & \numpm{1.741}{+0.069}{-0.074} & $1.753 \pm 0.036$ & $1.81 \pm 0.09$ & $1.878 \pm 0.042$ & \numpm{1.87}{+0.15}{-0.10} & $1.619 \pm 0.021$ \\
    $M_{\rm p}$ [$M_{\rm Jup}$] & \numpm{0.894}{+0.014}{-0.013} & $\leq 3.382$ &$1.157 \pm 0.070$ & $1.66 \pm 0.12$ & $1.83 \pm 0.47$ & $\leq 6.78$ & \numpm{1.99}{+0.16}{-0.14} \\
    log$g_{\rm p}$ [\si{\cm\per\g\squared}] & $2.809 \pm 0.037$ & $\leq 3.44$ & $2.97 \pm 0.03$ & $3.10 \pm 0.05 $ & $3.11 \pm 0.11$ & $\leq 3.68$ & $3.38 \pm 0.03$ \\
    $T_{\rm eq}$ [K] & $2160 \pm 40$ & $2262 \pm 73$ & $2358 \pm 52$ & $2470 \pm 60$ & $2492 \pm 38$ & \numpm{2562}{+43}{-52} & $2641 \pm 34$ \\
    \midrule
    References & \citet{ehrenreich_nightside_2020} & \citet{lund_kelt-20b_2017} & \cite{delrez_wasp-121_2016} & \citet{hellier_new_2019} & \citet{simonnin_time_2024} & \citet{Zhou_2019} & \citet{lendl_hot_2020} \\
    & \citet{West_2016} & & \citet{bourrier_hot_2020} & & \citet{cabot_toi-1518b_2021} &  & \citet{Anderson_2018}\\
    \bottomrule
    \end{tabular}}
    \textit{Notes:} The surface gravity log$g_{\rm p}$ of the planet is calculated based on the mass and radius, assuming Gaussian error propagation.
\label{tab:params}
\end{table}

\newpage
\section{Volume mixing ratios}

\begin{figure*}[h!]
    \centering
    \includegraphics[width=0.8\linewidth]{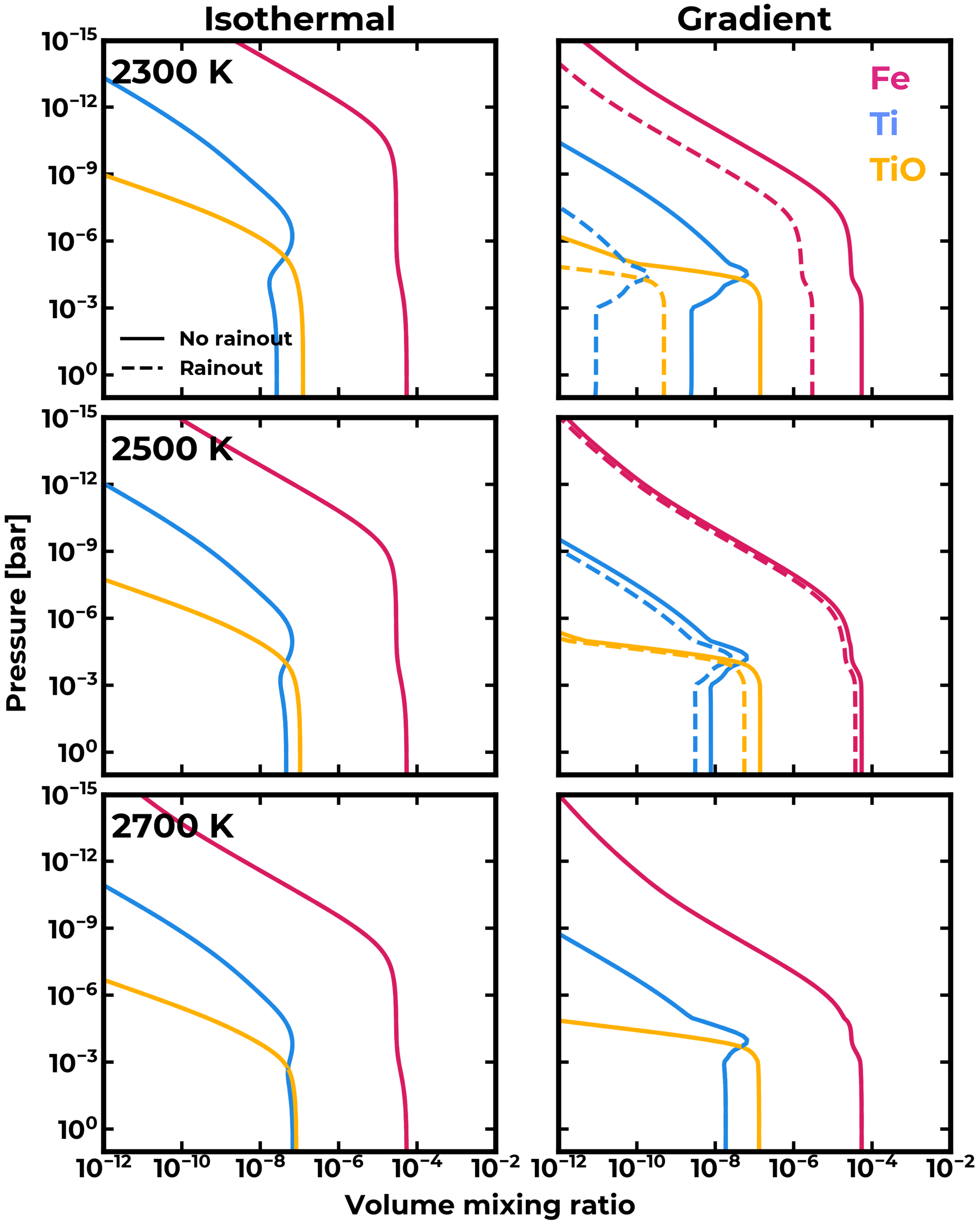}
    \caption{Volume mixing ratios of Fe (blue), Ti (pink), and TiO (yellow) for isothermal (left) and non-isothermal (right) atmospheres at \SI{2300}{\kelvin} (top), \SI{2500}{\kelvin} (middle), and \SI{2700}{\kelvin} (bottom). Solid and dashed lines indicate models without and with rainout, respectively. In the isothermal case, rainout is not effectively triggered despite being included, and the gas-phase Ti abundance is primarily governed by partitioning between atomic Ti, TiO, and ionised Ti. As a result, Ti remains distributed over a broad pressure range, leading to only modest changes in its observable signal. In contrast, the non-isothermal atmospheres develop cooler layers at depth where Ti is efficiently converted into TiO and subsequently removed through condensation (e.g. into \ch{CaTiO3}). This process depletes the available gas-phase Ti and confines the remaining atomic Ti to a narrower region at low pressures. While Fe also begins to condense at depth, it is less strongly affected over the pressures probed by the observations. Because the Ti signal is more sensitive to deeper atmospheric layers than Fe, this differential depletion leads to a pronounced suppression of the Ti signal relative to Fe, naturally explaining the strong decrease in \metric for atmospheres with temperature gradients.
    }
    \label{fig:partitioning}
\end{figure*}

\newpage
\section{Observations}

Table D.1: Overview of the observations considered in this study.
{\small
\begin{center}
\begin{longtable}{lllllllp{1cm}}

			\toprule
			{Planet }& {Date}  & {Instrument}    & {PI, \#}     & {S/N$^a$} & {T$_{eq}$ [K]} & {Atmospheric detections in $^b$} \\
             \midrule
			KELT-11  b       & 2024-03-31	& MAROON-X     & Parmentier, GN-2024A-Q-130         & 22.2-302.5  & 1712 & \\	
                             & 2024-04-19	& MAROON-X     & Parmentier, GN-2024A-Q-130         & 22.2-297.6  & \\
			\midrule
			WASP-172 b		 & 2022-06-01 & ESPRESSO & Albrecht, 109.22Z4.006 &  14.3-63.3 & 1740 & \citet{Seidel_Prinoth_2023} \\       
             \midrule
			KELT-4 A b & 2023-12-07	& MAROON-X  & Parmentier, GN-2023B-Q-127    &  28.0-82.7  & 1827 & \\	
            & 2023-12-10	& MAROON-X     & Parmentier, GN-2023B-Q-127         &  26.4-77.6   &  & \\
            & 2023-12-16	& MAROON-X     & Parmentier, GN-2023B-Q-127         &  24.7-76.3    &  & \\
            & 2023-12-31	& MAROON-X     & Parmentier, GN-2023B-Q-127         &   -   &  & \\
		    \midrule
			WASP-173A b &  2022-07-24 & ESPRESSO & Albrecht, 109.22Z4.003  & 12.4-54.9  & 1880 & \citet{Jiang2026}, non-detection\\
	
             \midrule
			KELT-7 b & 2023-12-28	& MAROON-X     & Parmentier, GN-2023B-Q-127   &  22.2-158.1 & 2048 & \\
			\midrule
			WASP-19 b 			& 2019-01-14 & ESPRESSO & Sedaghati, 0102.C-0311 &  13.2-60.7 & 2077 & \citet{sedaghati_spectral_2021}\\
		      & 2019-03-03 & ESPRESSO & Sedaghati, 0102.C-0311 &  11.8-56.1 && \citet{sedaghati_spectral_2021}\\
		      & 2019-03-22 & ESPRESSO & Sedaghati, 0102.C-0311 &  13.0-66.3 && \citet{sedaghati_spectral_2021}\\
			& 2020-01-11 & ESPRESSO & Sedaghati, 0102.C-0311 &   6.8-48.9 && \citet{sedaghati_spectral_2021}\\
			\midrule
			KELT-17 b	& 2020-12-25	& ESPRESSO & Seidel, 0106.C-0126   & 22.7-107.2 & 2087  & \\
							& 2021-01-31	& ESPRESSO & Seidel, 0106.C-0126 &  32.0-134.9 && \\
								& 2021-03-06	& ESPRESSO & Seidel, 0106.C-0126  & 25.5-111.4 & & \\
			\midrule
			WASP-76 b&2018-09-02   & ESPRESSO & Pepe, 1102.C-0744       & 13.1-83.7 &2228 & \citet{ehrenreich_nightside_2020}\\ 
			&2018-10-30   & ESPRESSO & Pepe, 1102.C-0744                 & 10.5-67.5&  & \citet{ehrenreich_nightside_2020}\\
            &2019-10-18   & ESPRESSO & Gibson, 0104.C-0642              & 29-55 && \citet{maguire_high_2024}\\
            &2020-09-03   & MAROON-X & Pelletier, GN-2020B-Q-122     &  20.0-68.4 && \citet{pelletier_vanadium_2023}\\
            &2020-09-12   & MAROON-X & Pelletier, GN-2020B-Q-122     &  17.6-63.1 && \citet{pelletier_vanadium_2023}\\
            &2021-10-28   & MAROON-X & Debras, GN-2021B-Q-138        &  29.4-99.0 && \citet{pelletier_vanadium_2023}\\
            \midrule
            KELT-20 b  & 2023-07-07 & MAROON-X &  Parmentier, GN-2023A-Q-224 & 71.3-190.6 & 2263 & \citet{de_lia_2026}\\
			\midrule
			WASP-121 b & 2021-01-26	 & ESPRESSO & Gibson,  0106.C-0516  &  27-37 & 2338 & \citet{maguire_high_2024}\\
			& 2021-03-04 & ESPRESSO & Gibson,  0106.C-0516 & 21-36  && \citet{maguire_high_2024}\\
           & 2019-01-06 & ESPRESSO & Pepe, 1102.C-0744 &  21-36  && \citet{borsa_atmospheric_2021} \\
           & 2018-11-30 & ESPR-4UT & 60.A-9128 (Comm.) &   23.1-163.3  & 2358 & \citet{borsa_atmospheric_2021} \\
           & 2023-09-23 & ESPR-4UT & Seidel, 111.24J8 &   26.6-174.5 && \citet{seidel_vertical_2025}\\
			\midrule
			MASCARA-4 b	  & 2020-02-12 & ESPRESSO & Wyttenbach, 0104.C-0605 300 & 44.9-201.7 & 2405& \citet{zhang_transmission_2022}\\
			& 2020-02-29 & ESPRESSO & Wyttenbach, 0104.C-0605  &  46.3-207.6 && \citet{zhang_transmission_2022}\\
            \midrule
            WASP-178 b & 2021-05-03 & ESPRESSO & Pepe, 1104.C-0350 &   11.3-48.6  & 2470 & \citet{damasceno_atmospheric_2024-1}\\
            & 2021-07-09 & ESPRESSO & Pepe, 1104.C-0350 & 9.3-37.7 &  & \citet{damasceno_atmospheric_2024-1}\\
             \midrule
			TOI-1518 b & 2022-08-13	& MAROON-X     & Parmentier, GN-2022B-Q-128     &23.4-64.9 &  2491 & \citet{simonnin_time_2024}\\
             & 2023-10-19	                & MAROON-X     & Parmentier, GN-2023B-Q-127     &  27.4-80.2 &  &   \citet{simonnin_time_2024}   \\
              & 2024-06-26	                & MAROON-X     & Parmentier, GN-2024A-Q-130   &  24.0-65.1 &  &  \citet{simonnin_time_2024}    \\ 
			\midrule
           HAT-P-70 b  &  2023-12-13 & 	MAROON-X & Pelletier, GN-2023B-Q-126 &  22.2-64.8 & 2559  &  \citet{Sun2026} \\
           &  2023-12-24 & 	MAROON-X & Pelletier, GN-2023B-Q-126 &  22.2-72.5 &  & \citet{Sun2026} \\
			\midrule
			WASP-189 b &2021-06-04   & ESPRESSO  & Prinoth, 107.22QF     &  107.7-432.6 & 2638 & \citet{prinoth_time-resolved_2023}\\
             &2022-04-03  & MAROON-X  & Pelletier, GN-2022A-FT-208      & 84.1-231.6 && \citet{prinoth_atlas_2024}\\ 
             &2022-06-02  & MAROON-X  & Pelletier, GN-2022A-FT-208     & 84.7-233.5 && \citet{prinoth_atlas_2024}\\ 
			\bottomrule
\label{tab:observation_log}
\end{longtable} \vspace{-2em}
\end{center}
\textit{Notes:} $^{(a)}$ Minimum and maximum S/N values across the full wavelength range, assuming the photon-noise limit. $^{(b)}$ First publication of the exoplanet's atmosphere from the specific dataset.
\end{document}